\documentclass[preprint,prl,showpacs,floatfix]{revtex4}

\usepackage{graphicx}

\begin{document}

\title
{
Discreteness-induced Transition in Catalytic Reaction Networks 
}
    
\author
{ 
Akinori Awazu
}

\affiliation
{
Department of Mathematical and Life Sciences, Hiroshima University,
Kagami-yama 1-3-1, Higashi-Hiroshima 739-8526, Japan.
}

\author
{ 
Kunihiko Kaneko
}

\affiliation
{
Department of Basic Science, University of Tokyo $\&$
ERATO Complex Systems Biology, JST,\\
Komaba, Meguro-ku, Tokyo 153-8902, Japan.
}

\date{\today}

\begin{abstract}
Drastic change in dynamics and statistics in a chemical reaction
system, induced by smallness in the molecule number, is
reported. Through stochastic simulations for random catalytic reaction
networks, transition to a novel state is observed with the decrease in
the total molecule number $N$, characterized by: i) large fluctuations
in chemical concentrations as a result of intermittent switching over
several states with extinction of some molecule species and ii) strong
deviation of time averaged distribution of chemical concentrations
from that expected in the continuum limit, i.e., $N \to \infty$.  The
origin of transition is explained by the deficiency of molecule
leading to termination of some reactions.  The critical number of
molecules for the transition is obtained as a function of the number
of molecules species $M$ and that of reaction paths $K$, while total reaction rates, scaled
properly, are shown to follow a universal form as a function of
$NK/M$.
\end{abstract}


\maketitle

\section{1. Introduction}

In intra-cellular biochemical reaction processes, some chemical
species often play an important role at extremely low concentrations,
amounting to only a few molecules per
cell\cite{cell1,cell2,cell3,cell4}. In such situations, the
fluctuations and discreteness in the molecule number are obviously
important.  On the other hand, in rate equations, generally adopted in
chemical kinetics, the concentration of each chemical species is
treated as a continuous variable, and the fluctuations and the
discreteness of the number of molecules are neglected. However, if the
molecule number is not very large, the number fluctuations as well as
discreteness in the number, rather than a continuous concentration,
has to be considered seriously.

Of course, effects of fluctuations in concentrations are considered by
using stochastic differential equations. Indeed, several non-trivial
noise-induced phenomena have been reported\cite{sto1,sto2,sto3}.
However, in most of such studies, discreteness in the molecule
numbers, i.e., the number being 0,1,2,.., has not been considered
seriously.

Recently, Togashi and Kaneko reported a drastic change in the steady
distribution of chemical concentrations as a result of discreteness in
the molecule number, by studying a catalytic reaction network with a few
molecules species\cite{togashi1,togashi2}. Novel types of dissipative
structure formation, induced by very low concentration molecules, have
also been investigated\cite{mif1,mif2,Solomon,togashi3,marion,zhdanov} 
in a class of reaction-diffusion systems or the models of 
biochemical reactions in cells. The observed novel states in
these studies are a result of fluctuations and discreteness in the
molecular numbers, in particular of extinction and re-emergence of
some molecule species, which alternate in time.

Relevance of such discreteness in molecule number to emergence of
novel states should not be restricted to a simple reaction network
with a few molecule species, but is also expected to exist in a wide
variety of chemical reaction systems with a large number of species.
Here, dynamics and statistics of chemical reaction systems with a
finite (small) number of molecules and a large number of molecule
species have to be investigated.  Such study is important not only for
biochemical reaction kinetics generally, but also as a problem of
non-equilibrium statistical mechanics. Here, the criterion on
``smallness'' in the number itself should be clarified as a condition
for discreteness-induced transition for a given chemical reaction
network system.

In this paper, we study discreteness-induced transition in a reaction
network where a large number of chemical species is connected by
catalytic reaction paths chosen randomly.  Use of random catalytic
reaction network is pioneered by Kauffman \cite{Kauffman,Kaneko-Adv}
for the problem of the origin of life, while studies in a growing cell model
consisting of such reaction network have 
unveiled universal statistical behaviors of chemical concentrations,
which are confirmed in the gene expression data in the present
cells\cite{Furusawa,Ito}. Here, we are interested in how discreteness
in the molecule numbers affect global behavior of chemical reaction
dynamics in such network. For simplicity, we only consider the
reaction network whose steady state is unique and stationary when the
number of molecules is infinite, i.e., the corresponding rate equation
has just a unique fixed point attractor. Even in such a simple system,
we find the following transition to a discreteness-induced state,
which appears when the total number of the molecules is below a
critical value;

A) Chemical concentrations exhibit intermittent switching among
several states with distinct chemical compositions.

B) The long time average of the chemical concentrations deviate
distinctly from those expected in the continuum limit with a large
number of molecules where the rate equation description is valid.

We will also obtain the critical molecule number for this
discreteness induced transition, whose dependence on the number of
chemical species and path ratio of the reaction network will be
derived.

In the next section, we introduce a specific reaction network model,
while numerical results to show the above transition with the decrease
in the molecule number are presented in section 3. Possible mechanism
for the transition is described in section 4, where deficiency in some
molecules is shown to introduce switching among several effective
reaction networks that consist only of non-vanishing chemical species.
In section 5, we obtain the critical number of molecules for the
transition as a function of the number of molecule species and
connectivity in the network.  Summary and discussion will be given in
section 6.

\section{2. Model}

Now, we introduce a simple model of a network of elementary
reactions that consists of a variety of chemical
species\cite{Furusawa}. State of the system is represented by a
set of numbers $(n_1, n_2, .... ,n_M)$, where $n_i(=0,1,...)$ 
indicates the number of molecules of the chemical species 
$i$($1 \le i \le M$), with $M$ 
as a total number of molecule species.  Here, the total
number of molecules is fixed at $N$, and accordingly $0 \le
n_i \le N$. For the chemical reaction dynamics, we choose a catalytic
network among these chemical species, where each reaction from a
chemical $B$ to another chemical $A$ is assumed to be catalyzed by a
third chemical $C$; i.e., $B + C \to A + C$ (see Fig. 1).  The
reaction coefficients are set to be identical for all reactions for
simplicity, and chosen to be $1$. 
Then, the growth rate in $n_A$ (or the decay rate in $n_B$) through this
reaction is give by $n_B n_C/N^2$, on the average.

The connection paths in a reaction network are chosen randomly (and
then fixed), where the average number of the reaction paths from a
chemical $i$ to any other chemical species $j$ catalyzed by a chemical
$l$ is set at a given connection number $K$. We do not include
auto-catalytic reaction in the form of $B + C \to 2C$, because such
type of reaction is not usually elementary but is realized as a result
of a series of (non-auto-catalytic) elementary reactions. Also,
inclusion of auto-catalytic paths sometimes leads to non-fixed-point
or multiple attractors, which makes discussion on the discreteness
effect complicated.

We also include a flow of chemicals into and out of the system from
the reservoir. With this process, the total number of molecules is fixed so that a
molecule is regarded to be replaced by some other with a certain rate.  Instead of
considering such flow, one can equivalently consider a combination of
decomposition and synthesis of some chemicals, or non-catalytic
changes between different molecule species that are chosen randomly
from all chemicals with equal probability.  For simplicity, we assume
that this `non-catalytic' change occurs with equal probability
$\epsilon$ for all molecules, while its rate is much smaller than that
of the catalytic reaction.

Numerical simulations are carried out by iterating the following
stochastic processes. First, we randomly pick up a pair of molecules
and if the pair is a substrate and catalyst according to the reaction
network, the substrate molecule is transformed to the product molecule
according to the reaction. Second, we randomly pick up a molecule and
transform it by a non-catalytic change, with a given, much lower, rate. 
Here, a unit time is given as the time span in which Monte-Carlo steps 
for catalytic reactions are repeated $N/2$ times and those for 
non-catalytic processes are repeated $N$ times. In each time, each 
molecule collides with another molecule once on average, to check if the 
catalytic reaction occurs, while it is transformed transformed to some 
other chemical with probability $\epsilon$ by non-catalytic process.
Numerically, we apply this stochastic simulation, while in the cases
with $N >> M$ (i.e., continuous limit), the reaction dynamics are
represented by the following rate equation,
\begin{equation}
\dot{n_{i}} = \sum_{j,l} C(j, i, l) \frac{n_{j}n_{l}}{N^2} - \sum_{j',l'} C(i,
 j', l') \frac{n_{i}n_{l'}}{N^2} + \epsilon(\frac{1}{M} - \frac{n_{i}}{N}),
\end{equation}
where $C(i, j, l)$ is 1 if there is a reaction $i + l \to j + l$, and 0
otherwise. 

In the following sections, we present numerical results of the
stochastic simulations and show how the steady state properties of
such catalytic reaction networks depend on the number of chemical
species $M$ and of molecules $N$. In this paper, we consider the case
with $K$ as $K_c < K << M$ where $K_c$ indicates the critical
connection number of the percolation transition in random networks. It
is noted that the rate equation (1) in the continuum limit has only a
unique fixed point attractor that gives the concentrations at the
steady state.

\section{3. Discreteness-induced Transition}

We show the results of the stochastic simulation of catalytic reaction
networks with $\epsilon = 10^{-4}$. We study how dynamical aspects of
the system change depending on the number of molecules $N$, by taking
a given reaction network. Figures 2(a) and (b) show typical temporal
evolutions of the concentrations of each chemical species $n_i$ for a
reaction network with $M = 100$ and $K = 12$. Two cases with
(a) large ($N = 800$) and (b) small ($N = 12$) $N$ are plotted, while
Figure 2(c) shows the reaction ratio ($RR$) of these cases, where $RR$
is defined by [Number of reacting molecules]$/N$ at each time.  If $N$
is much larger than $M$ (e.g., $N = 800$), as seen in Figs. 2(a) and
(c), the distribution of $n_i$ is almost stationary and $RR$ takes a
constant value except for small fluctuations.

On the other hand, if $N$ is much smaller than some value (that is of
the order of $M$) (e.g., $N=12$), the distribution of $n_i$ exhibits
remarkable changes in time, and $RR$ fluctuates intermittently between
large values and $0$, as seen in Figs. 2(b) and (c).  Such
non-stationary behavior is caused by the discreteness in the molecular
number. When $N$ is much smaller than some threshold of the order of
$M$, the concentrations of some chemical species $n_i$ go to $0$ at
some time instance. Then the number of reactive chemicals or of the
catalysts often goes to $0$, for all reaction paths. Then, the
catalytic reactions of the system freeze, while the system can escape
from such freezing state as a result of non-catalytic changes (flow of
molecules).  Thus, $RR$ changes intermittently with time.

Next, we focus on typical statistical aspects of the system by changing
$N$. Long time average of the distributions of the chemical concentrations 
$<n_i/N>$ for a typical reaction network with $M=100$ and $K=12$ is plotted 
in Fig. 3(a) (for large $N$, $N = 800$) and Fig. 3(b) (for a small $N$, 
$N =12$). As seen in these figures, the
profiles of the average distribution are quite different between the two.
In Figs. 3(c) and (d), we show $<n_i/N>$ and the rank of abundances of
each chemical concentration, which is labeled in the order of magnitude of
 $<n_i/N>$. These are plotted as a function of $N$ for a given
reaction network, where each successive curve indicates the change in
$<n_i/N>$ and the rank of each chemical concentration.

The results in Figs. 3(c) and (d) suggest the existence of a critical
value of $N$, denoted by $N_c$. For $N$ smaller than $N_c$, the
chemical abundances of each species or their rank changes sensitively
with $N$. The variance of $<n_i/N>$ over time takes a maximum at some
$N$ slightly smaller than this critical value $N_c$. On the other
hand, for $N$ larger than $N_c$, $<n_i/N>$ are almost constant except
for the small fluctuations, and indeed the profile of $<n_i/N>$ for
such larger $N$ agrees with that obtained by the rate equation (1),
i.e., the system is well described by the continuum limit $N \to
\infty$.

Now, we focus on the transition of reaction dynamics at $N_c$.  For
$N$ larger than $N_c$, the temporal evolution of the distribution of
$n_i$ is almost stationary except for small fluctuations as in
Fig. 2(a). On the other hand, for $N < N_c$, there appears
intermittent switching between different distributions of $n_i$ as
shown in Fig. 2(b). In this case, the total reaction ratio is much
smaller and the temporal fluctuation of each chemical concentration is
much larger than the case with $N > N_c$. We have computed $N$
dependence of the average reaction ratio $<RR>$ and the average
temporal fluctuation (ATF) of all chemical concentrations.  Here,
$<RR>$ is defined as the long time average of $RR$, and ATF is defined
as the long time average of $\frac{1}{N}\sum_i (n_i - <n_i>)^2$.

Figure 3(e) shows $<RR>$ and ATF/($N/M$), plotted as a function of $N$
for the same reaction network. In this figure, $<RR>$ starts to
decrease drastically with the decrease in $N$, at the above mentioned
critical value. This figure also shows that ATF is proportional to $N$
for $N > N_c$, while ATF/($N/M$) increases sharply with the decrease in
$N$ for $N < N_c$.  This increase in the fluctuation is consistent with
the fact that the switching behavior becomes dominant for $N < N_c$.

As shown in Fig. 1, switching over several states with different 
effective network occurs for small $N$, as will be also discussed in 
the next section. As $N$ is decreased further below $N_c$, the 
reaction occurs rarely, and thus the frequency of such 
switching decreases, while for large $N$, the dynamics exhibit only 
small fluctuations around a stationary state. Thus, the fluctuation 
is expected to have a peak around $N = N_c$ where the 
switching occurs most frequently.

\section{4. Effects of molecular deficiency in a small reaction network}

The behavior observed in the previous section is rather common over a
variety of catalytic reaction networks. The dynamics and
the time averaged distributions of $n_i/N$ for a case with $N
<< N_c$ differ distinctly from those with $N \to \infty$, because of
the discreteness in the number of molecules. Here we discuss the
mechanism of this transition, by taking a simple example of the
catalytic reaction network with a small number of chemical
species. This example system is a little specific but can illustrate
the changes in the steady distribution and the effective network
structure consisting only of non-vanishing chemical components.  (see
\cite{togashi4} for a discreteness-induced switching over states in an
autocatalytic network).

We consider a network of catalytic reactions displayed in Fig.4 (a).
Here, we assume that $n_S > 0$ always holds.  By straightforward
calculation, the rate equation for the chemical concentrations has a
unique fixed-point attractor, which satisfies the relations, 
$2<n_A> = 2<n_B> = <n_C> = <n_D> = 2<n_E> = 2<n_R>$.

On the other hand, for small $N$, some of $n_i$ often happens to be
$0$.  In such cases, the above relationship on the fixed
concentrations no longer holds, where the distribution of the chemical
concentrations changes temporally among some characteristic
distributions with different relations among $<n_A>$, $<n_B>$,
$<n_C>$, $<n_D>$, $<n_E>$ and $<n_R>$.  

I) State with $n_A =0$ and $n_C =0$: When $n_A$ happens to be $0$,
$n_C$ also goes to $0$, if the decrease in $n_C$ by the reaction $C+E
\to $ progresses before the reaction $R+C \to A+C$ takes place.  This
state is also reached when $n_C$ happens to be $0$ and if $n_A$ goes to
$0$ by the reaction $A + R \to D + R$ or by the non-catalytic
process. Once both $n_A $ and $n_C $ vanish, this state with
extinction of both chemical species is preserved over a long time,
because neither $A$ nor $B$ is synthesized by catalytic reactions, and
only slow non-catalytic changes can produce such chemicals.

At this state, none of the reactions catalyzed by $A$ or $C$ take
place. Then, as long as $n_i >0$ for $i \ne A$, $C$, the effective
structure of the reaction network is reduced to that shown in
Fig. 4(b). Only molecules within this sub-network exist.  (In
Fig. 4(b), the arrows indicating the reactions catalyzed by $A$ or $C$
are removed.)  In this case, $n_D$ increases, because the reaction to
decrease $n_D$ catalyzed by $A$ does not take place. Accordingly, the
reaction $S+D \to T+D$ takes place frequently so that the reaction
through the series $S \to T \to G \to H \to$ progresses with a high
rate.

II) The state with $n_G =0$ and $n_H =0$: When $n_G$ happens to be
$0$, $n_H$ also goes to $0$, if $n_H$ decreases by the reaction 
$H+D \to $. 
It also appears when $n_H$ happens to be $0$ and $n_G$ decreases to $0$
by the reaction $G + C \to F + C$. Once both $n_G $ and $n_H $ vanish,
this state is preserved over a long time.

Here, when $n_i >0$ for $i \ne G$, $H$, the effective structure of the
reaction network is given in Fig. 4(c).  Similarly with the case I),
chemicals are localized within this subnetwork.  In this effective
network the reaction progresses through the series $S \to R \to A
\to$, $S \to R \to B \to$ and $S \to R \to E \to$ with a high rate.

III) For $i \ne$ $A$, $C$, $G$ and $H$, $n_i$ is soon recovered even
if it happens to be 0. For example, even if $n_B$ happens to be $0$,
as long as $n_D$ does not reach 0, $n_B$ is soon recovered by the
reaction $R+D \to B+D$. Other chemical species also behave in a
similar manner. Then, disconnection of the reaction paths does not
occur.

Each of these three states has a long life time when the number of
molecules $N$ is much smaller than the number of the chemical species.
Stochastic switching over such states is commonly observed for a
reaction network system with small $N$, as given in Fig. 2(b).  Due to
these switchings, the temporal fluctuation of each chemical
concentration is enhanced, as given in Fig. 3(e). Thus, the behavior
for small $N$ is distinct from that obtained in the case with $N \to
\infty$.

Here the relations between $<n_A>$, $<n_B>$, $<n_C>$ and $<n_D>$
differ distinctly from that in the continuum limit.  For example,
$<n_A> < <n_B>$ and $<n_C> < <n_D>$ are obtained here.  This deviation
from the continuum limit is understood easily by considering the
number distribution at the states I) and II).  

In general, we expect that situations similar to this simple example
should appear in some part of the reaction networks for a system with
a large number of the chemical species, as discussed in the last
section. This leads to the transition at $N < N_c$, with drastic
difference in the number distribution from the continuum limit, as
given in Fig. 3(a), (b) and (c).

\section{5. Critical value of molecular number}

In section 3, we suggested the existence of the critical total number
of molecules $N_c$, at which the behaviors of random catalytic 
networks change drastically.  For $N > N_c$, the behavior is well
represented by continuum description, while for $N < N_c$, deficiency
in some molecule species suppresses the ongoing reaction. In this
section, we study dependence of this critical number $N_c$ on $K$ and
$M$ quantitatively, by computing the reaction rate $RR$.

Figure 5 shows examples of dependence of $<RR>$ on $N$ for several
values of $K$, $K=8, 12, 16, 20$ and $24$, and $M$, (a) $M=100$
with $\epsilon = 10^{-4}$, (b) $M=1000$ with $\epsilon = 10^{-4}$.
As $N$ is decreased, there is a drastic drop in $<RR>$ at some value
of $N$.  When $N$ is sufficiently larger than this value, $<RR>$ is
almost constant, approaching a value at the continuum limit 
$\sim O(<RR>_{N \to \infty})$ where chemical concentrations are almost
stationary except for small fluctuations. On the other hand, for $N$
smaller than this critical value, the chemical concentrations exhibit
intermittent switching among several states. 

Now, we consider $<RR>$ for a specific case with $N \to \infty$; 
for all chemical species, the number of the reaction path from
other chemical species and those to other chemical species are given
a unique value $K$. In such cases, the distribution of the chemical
concentrations goes to uniform. In each time step, each molecule collides 
one molecule on average, and the probability that the molecule catalyzes 
the reaction from the collided molecule is $K/M$. Then,
$<RR>$ is given by $K/M$. 

If the fluctuation of the number of the reaction paths for each chemical 
specie increases, the reaction from the chemical species which have small
number of reaction paths limit the reaction rate.
Then, $<RR>$ is smaller than $K/M$ in general.

Now, to study dependence of $<RR>$ on $K$, $M$, and $N$, we 
introduce the following scaling functions; $\rho = <RR> / (\frac{K}{M})$ 
as a normalized reaction rate and $N K / M$. Note that $N/M$ is nothing
but the average molecular number of
each chemical, which gives the average probability that each reaction
path is catalyzed. Thus, $N K / M$ gives the average number of the
effective reaction paths in the network. Based on these
considerations, we plot this normalized reaction rate as a function of
$NK/M$, to see $N,K,M$ dependence.

Figure 6 shows $\rho$ as a function of $N K / M$ for several values
of $K$, $K=8, 12, 16, 20$ and $24$, and $M$, (a) $M=100$, (b)
$M=1000$ and (c) $M=3000$ with $\epsilon = 10^{-4}$.  As $K$ is
increased, this scaled function approaches a form independent of $K$.
Here we note that there is a specific value of $N K / M = \kappa_c$ at
which all the curves of $\rho$ for different values of $K$ crosses.
Below this value $\kappa_c$, the normalized reaction rate decreases with
the increase in $K$, suggesting that the deficiency in molecule number 
per reaction path further suppresses the ongoing reaction. On the other
hand beyond $\kappa_c$, the normalized reaction rate slightly decreases 
with the decrease in $K$, due to the increase in the reaction path number
fluctuation per path. 
As the reaction is suppressed due to deficiency in molecule number for
$NK/M < \kappa_c$, this value of $\kappa_c$ gives a criterion for
discreteness induced transition.

Dependence of the reaction rate on $\epsilon$ is shown in Fig. 7. Here,
the value of $\kappa_c$ seems to be independent of $M$ and $\epsilon$
for larger $M$ ($M=1000$, $M=3000$ or larger $M$) and smaller $\epsilon$
(as is also compared with that in Fig.6), where $\kappa_c \sim 0.8$ holds.
On the other hand, the value of $\rho$ at $N K / M = \kappa_c$ decreases
as $\epsilon \to 0$, as shown in Fig. 6(c) and Fig. 7. Then, for
$NK/M<\kappa_c$, the reaction rate $\rho$ approaches 0, while $\rho$ at
$N K / M > \kappa_c$ is almost unchanged. In other words, as $\epsilon$
goes to zero, the normalized reaction rate seems to approach a step
function with 0 for $NK/M<\kappa_c$. This means that the catalytic
reactions often freeze.  Some reaction paths are terminated frequently
if $NK/M < \kappa_c$, and only non-catalytic changes give dominant
contributions to the reaction dynamics. On the other hand, for 
$N K / M > \kappa_c$, all chemicals can react along the connected paths in the
catalytic reaction network for most of time, so that the behavior at the
continuum limit is valid, i.e., $<RR> \sim O(<RR>^K_{N \to \infty})$.
To sum up the value $\kappa_c$ gives a criterion for the
discreteness-induced transition, i.e., $N_c \sim \kappa_c \times M / K$
for large $M$.

\section{6. Summary and discussions}

In this paper, we have reported a discreteness-induced transition in
catalytic reaction networks with random connections.  When the total
number of the molecules is smaller than a critical value, transition
to a novel dynamical state is observed with a distinct behavior from
that expected in the continuum limit, i.e., the molecular number $\to
\infty$. The behavior is characterized by switching over
quasi-stationary states where some reaction paths are terminated
effectively due to deficiency in molecule numbers.  Each
quasi-stationary state is characterized by a smaller set of effective
reaction networks.

The critical molecule number for this transition is shown to be
proportional to the number of chemical species $M$ divided by the
average number of reaction path per species $K$, with a proportion
coefficient, estimated to be $\kappa_c \sim 0.8$ in a limit 
with large $M$. Whether the number of molecules is large enough to be
approximated by the continuous rate equation or not is thus determined
by $N ^{>}_{\sim} 0.8M/K$.

So far we have not succeeded in estimating this value of $\kappa_c$
analytically. It could possibly be related with the percolation
threshold of a random network, although the relationship 
is not so straightforward.  The percolation transition
point $k_c$ in general random networks is known to increase
logarithmically with the increase in the number of nodes of the
network\cite{graph} while $\kappa_c$ decreases to converge to a constant
value, with the increase in the number of nodes (chemical species).
Thus, the relation between $k_c$ and $\kappa_c$ is still unclear,
and analytic estimate of $\kappa_c$ is still an open question.

It is noted that the random network does not give a good approximation 
of the real biological network. However, the results obtained in such 
simplest reaction network should give a base to characterize the 
behaviors in several types of networks. The present study gives a 
starting point for the statistical physics of several catalytic reaction 
networks. 

On the other hand even by random catalytic reaction networks, some 
universal features such as Zipf's law in the gene expressions and log 
normal distributions of the fluctuations of chemical concentration in 
cells are reproduced\cite{Furusawa,Ito}. 
We may expect our results provide some insight to biological phenomena.   

In the presented model, below the critical number of molecules (or
beyond the critical number of molecule species), there appears
intermittent transitions over several states, which leads to the change
in the structure of effective reaction network. Such dynamics may give a
hint to uncover possible mechanisms of switching behavior in the
signaling pathways\cite{sig}. 

Also, dominantly acting paths in the metabolic reaction network of E. coli 
change in accordance with the concentration of the nutrient, as observed by the 
flux analysis\cite{meta}. 
In the nutrient poor environment, some chemical species may be deficient 
in a cell. The discreteness induced transition studied here may be relevant 
to such switch in the effective network. 

Recently, large phenotype fluctuations in isogenetic cells are reported 
in several organisms\cite{Ito,Laha,Miha}. 
In such cells, abundances of some molecule species are not so large 
(10 - 100 molecules) while there exist
more than 1000 chemical species\cite{Gupt}.
The intermittent switching induced by molecular deficiency may underlie 
such large fluctuations.

Last, we briefly discuss dependences of the transition on the network 
topology or a connecting distribution. 
We have confirmed that the value of $\kappa_c$ as well as the peak (or dip) 
in $<n_i/N>$ and ATF are almost unchanged, even if the reaction coefficients 
are distributed uniformly between $0$ ands $1$, or take two distinct values, 
say  0.1 and 0.9 randomly. 
The results are not changed either, even if there are some chemical species 
that catalyze many reaction paths and those catalyzing only a few paths.

However, the value of $\kappa_c$ can be shifted if the auto-catalytic 
productions are dominant. If the reaction to a chemical
species is catalyzed by itself, its concentration can increase rapidly,
which may decrease the number of the species at the upstream of
this reaction towards zero. The probability that some molecule species 
goes extinct is increased, as has been recently reported for a
simple reaction network of a few species \cite{togashi1,togashi2}.

Dependence of the discreteness-induced
transition on the network topology is an important future
issue both for chemical reaction network dynamics in general and also for 
the understanding of intra-cellular chemical reactions.

\section{ACKNOWLEDGMENT}
The authors would like to thank Y. Togashi, K. Fujimoto, M. Tachikawa 
and S. Ishihara for stimulating discussions.

\newpage

\begin{figure}
\begin{center}
%\psbox[width=8.0cm]{}
\includegraphics[width=8.0cm]{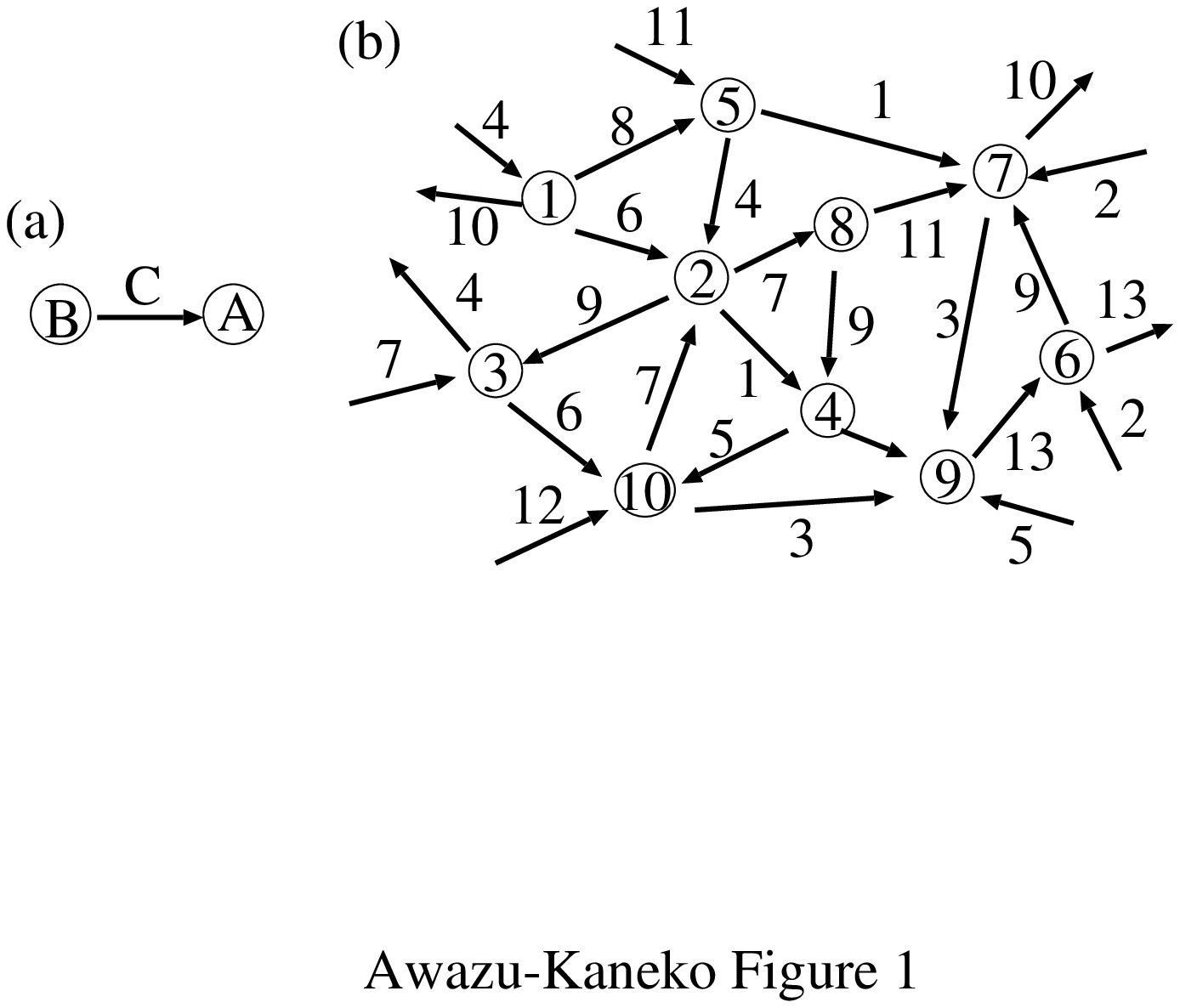}
\end{center}
\caption{(a) Illustration of catalytic reaction $B+C \to A+C$, and (b)
an example of catalytic reaction networks.}
\end{figure}

\begin{figure}
\begin{center}
%\psbox[width=8.0cm]{}
\includegraphics[width=7.0cm]{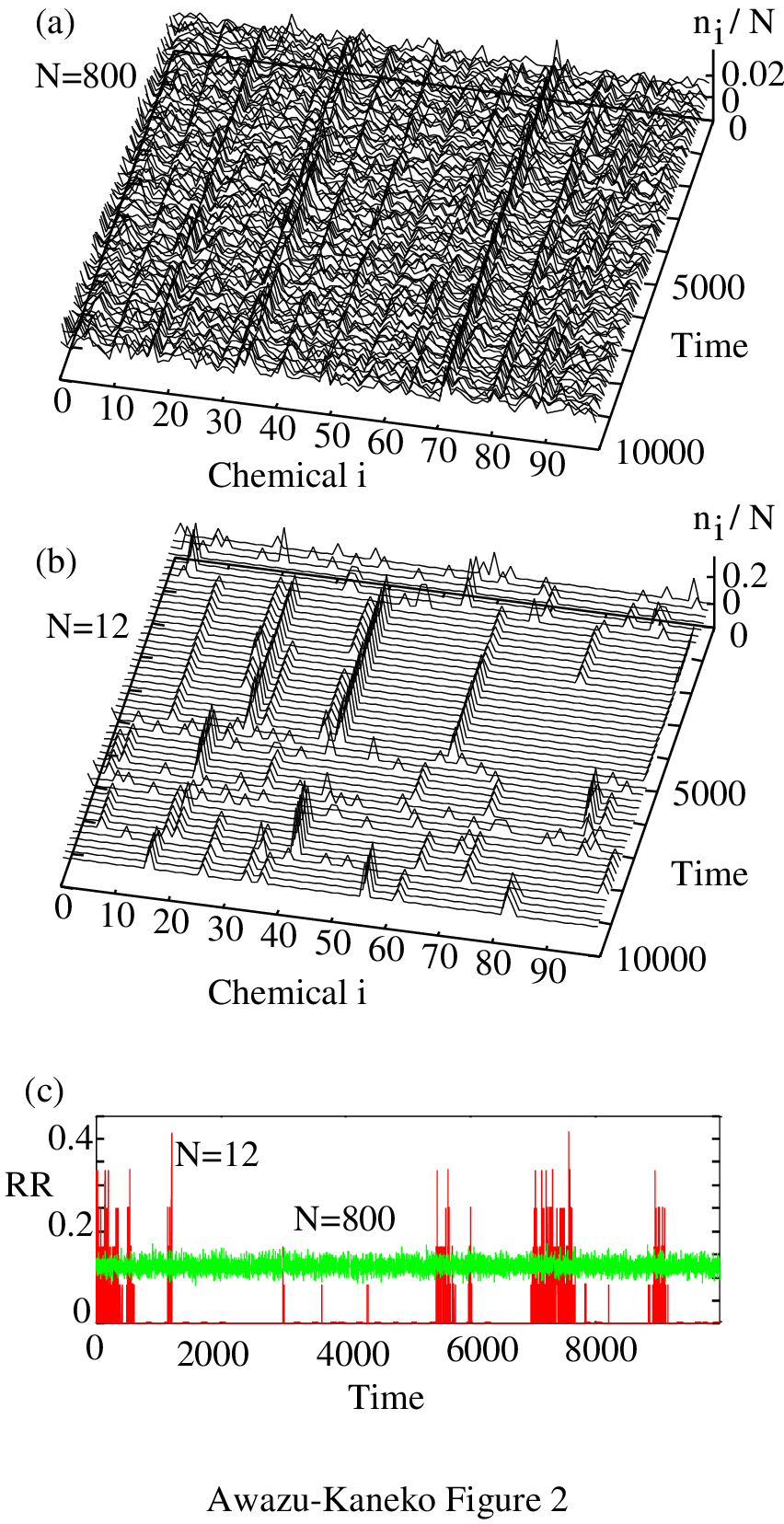}
\end{center}
\caption{Examples of temporal evolutions of $n_i$ for a given reaction
network for the case $M=100$ and $K = 12$ with (a) $N = 800$ and (b)
$N =12$. (c) Temporal evolutions of $RR$ for $N = 800$ (green) and $N
=12$ (red). The network is chosen randomly, but the behavior here is
typical, and observed over most networks generated randomly.}
\end{figure}

\begin{figure}
\begin{center}
%\psbox[width=8.0cm]{t}
\includegraphics[width=8.0cm]{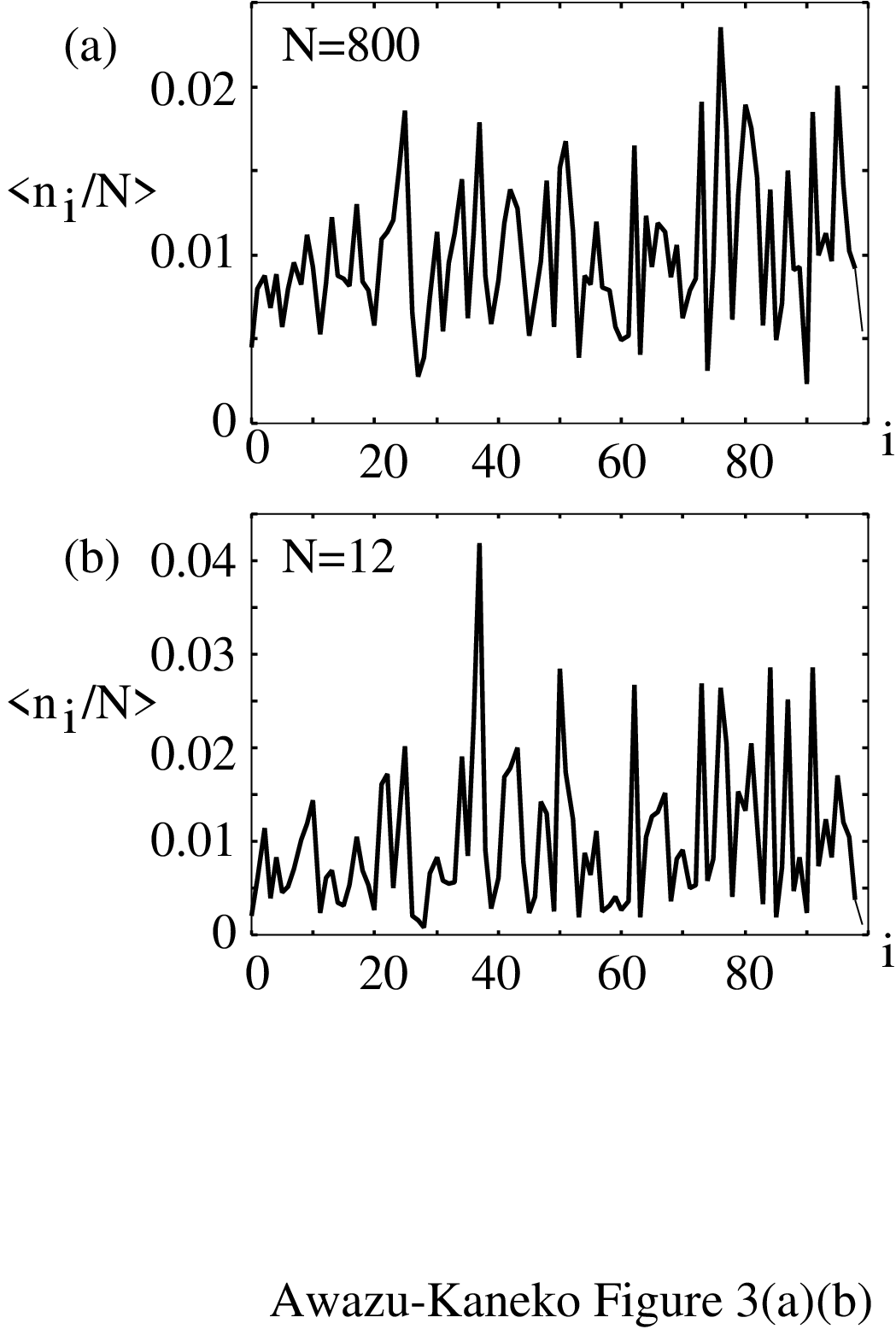}
\includegraphics[width=8.0cm]{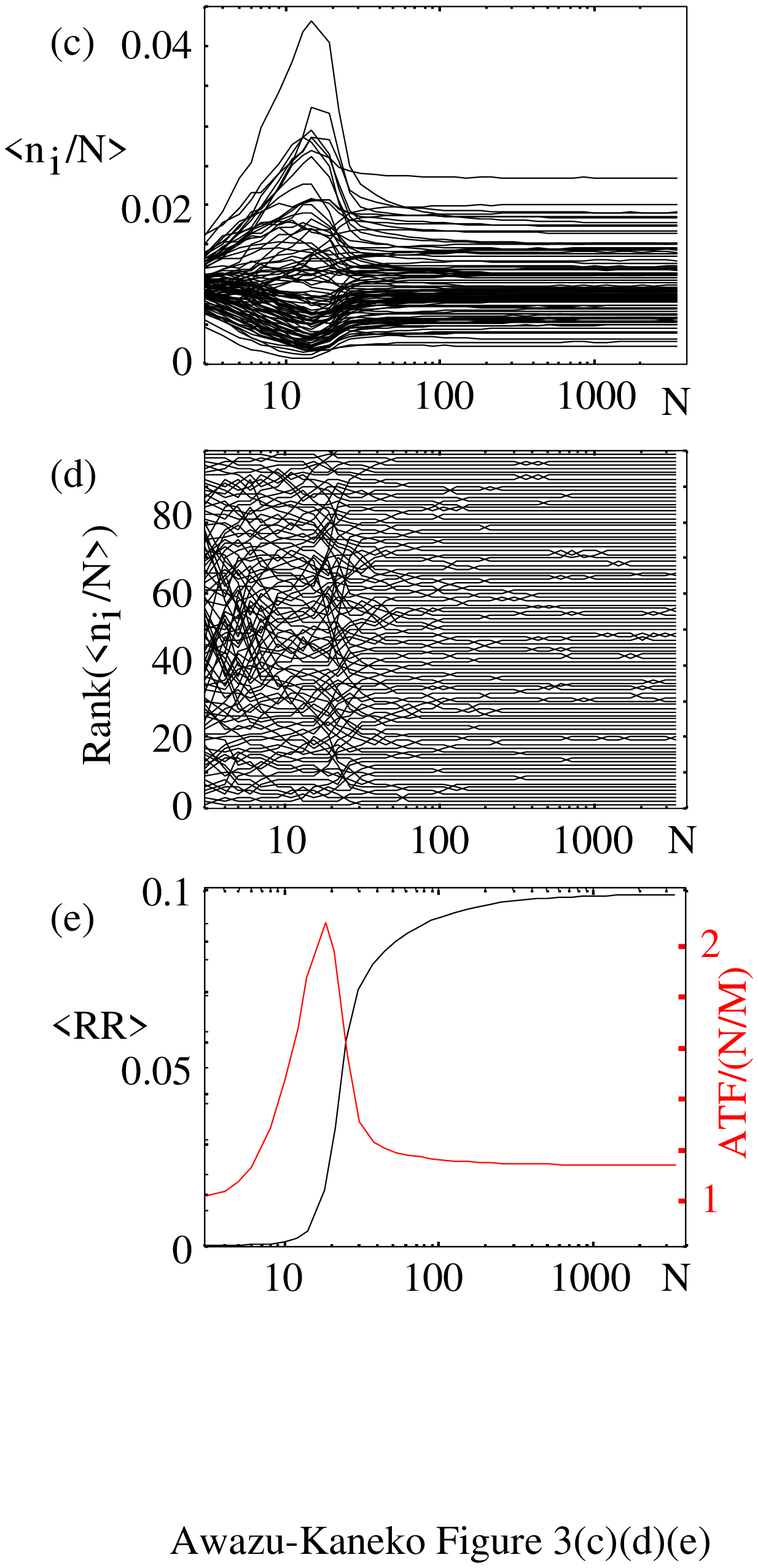}
\end{center}
\caption{(a)(b) $<n_i/N>$ of the same system as shown in the previous
 figure with (a) $N = 800$ and (b) $N =12$. (c)(d)(e) $N$ dependency
 of (c)$<n_i/N>$, (d) the rank of each chemical concentration, and (e)
 $<RR>$ and ATF/(N/M)).}
\end{figure}

\begin{figure}
\begin{center}
%\psbox[width=8.0cm]{}
\includegraphics[width=6.0cm]{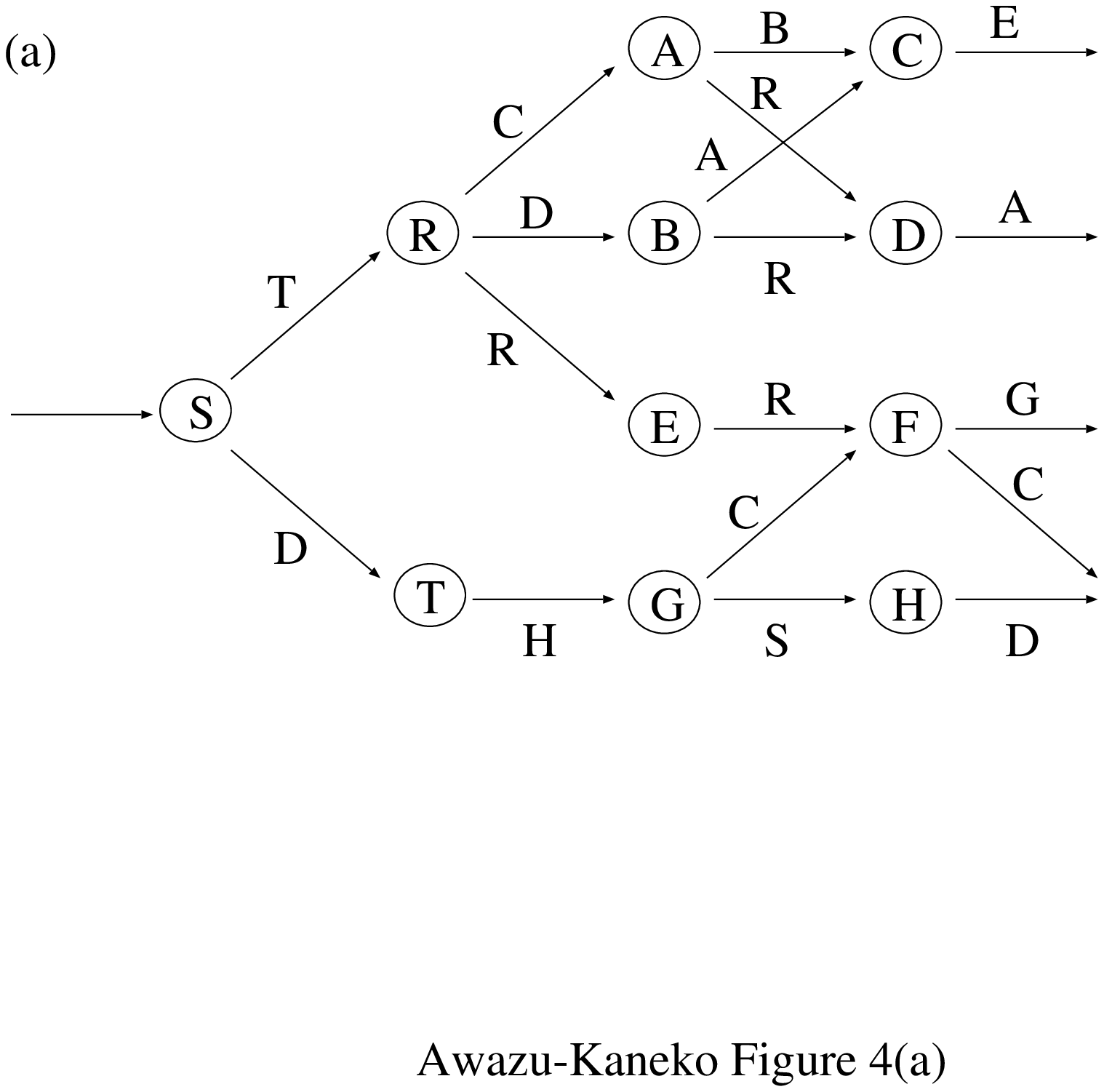}
\vspace{5mm}

\includegraphics[width=6.0cm]{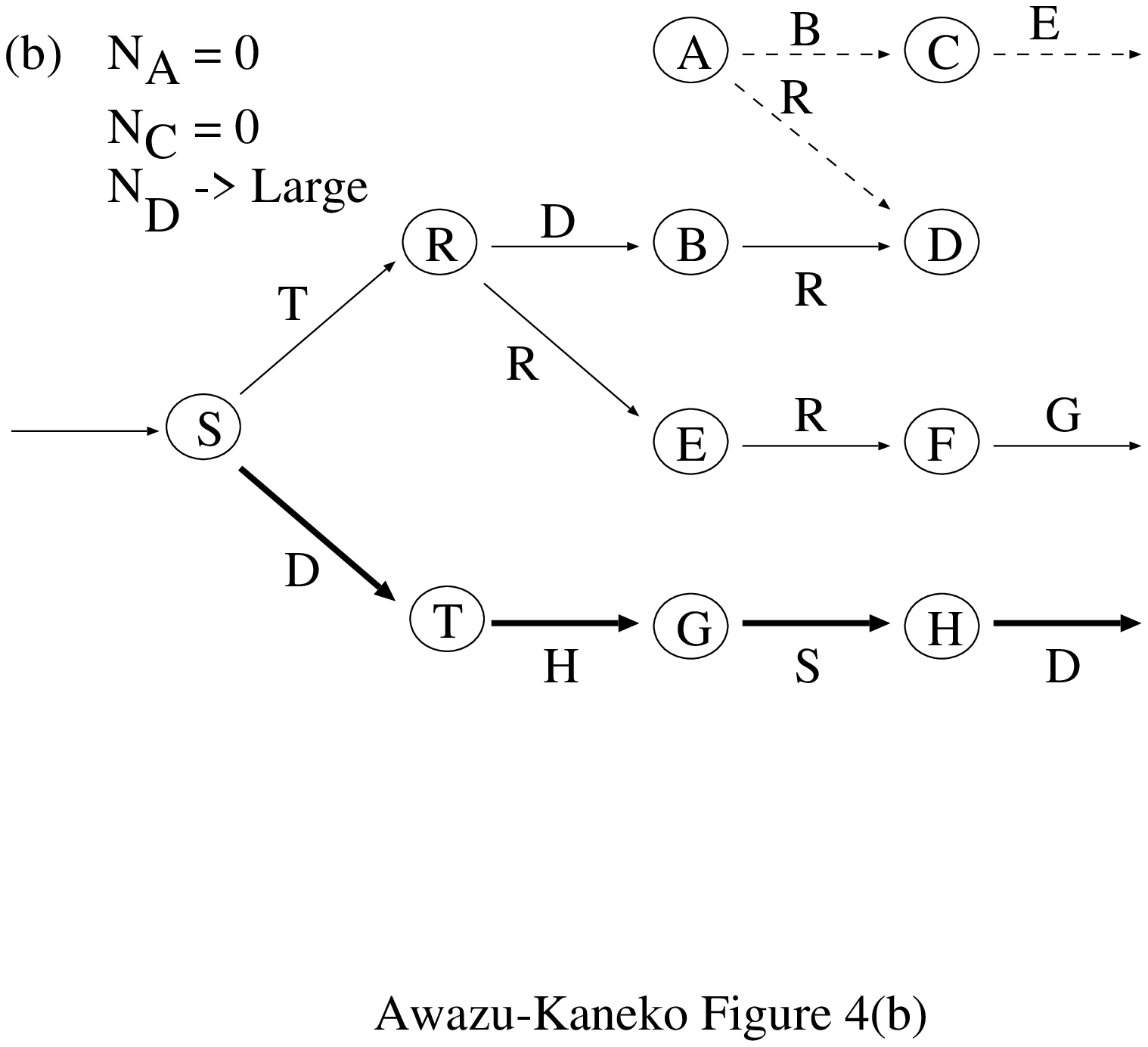}
\vspace{5mm}

\includegraphics[width=6.0cm]{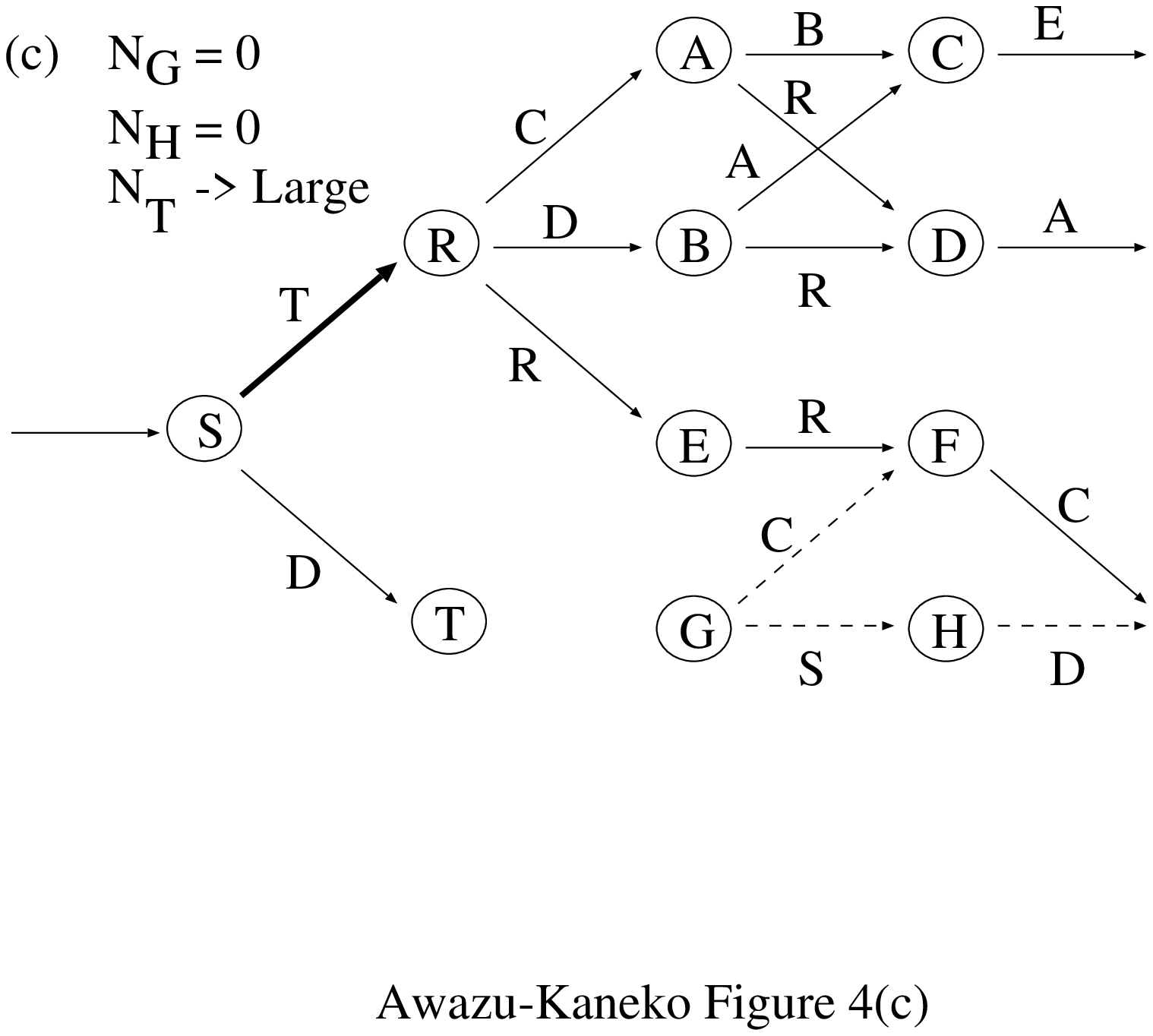}
\end{center}
\caption{Simple example of catalytic reaction network (a), and
 effective reaction network consisting only of non-vanishing chemicals
 for the state I) (b) and II) (c) described in the text. Thick
 and dashed arrows indicate the paths with a high reaction rate and
 those from the extinct chemical species at the moment, respectively.}
\end{figure}

\begin{figure}
\begin{center}
%\psbox[width=8.0cm]{}
\includegraphics[width=7.0cm]{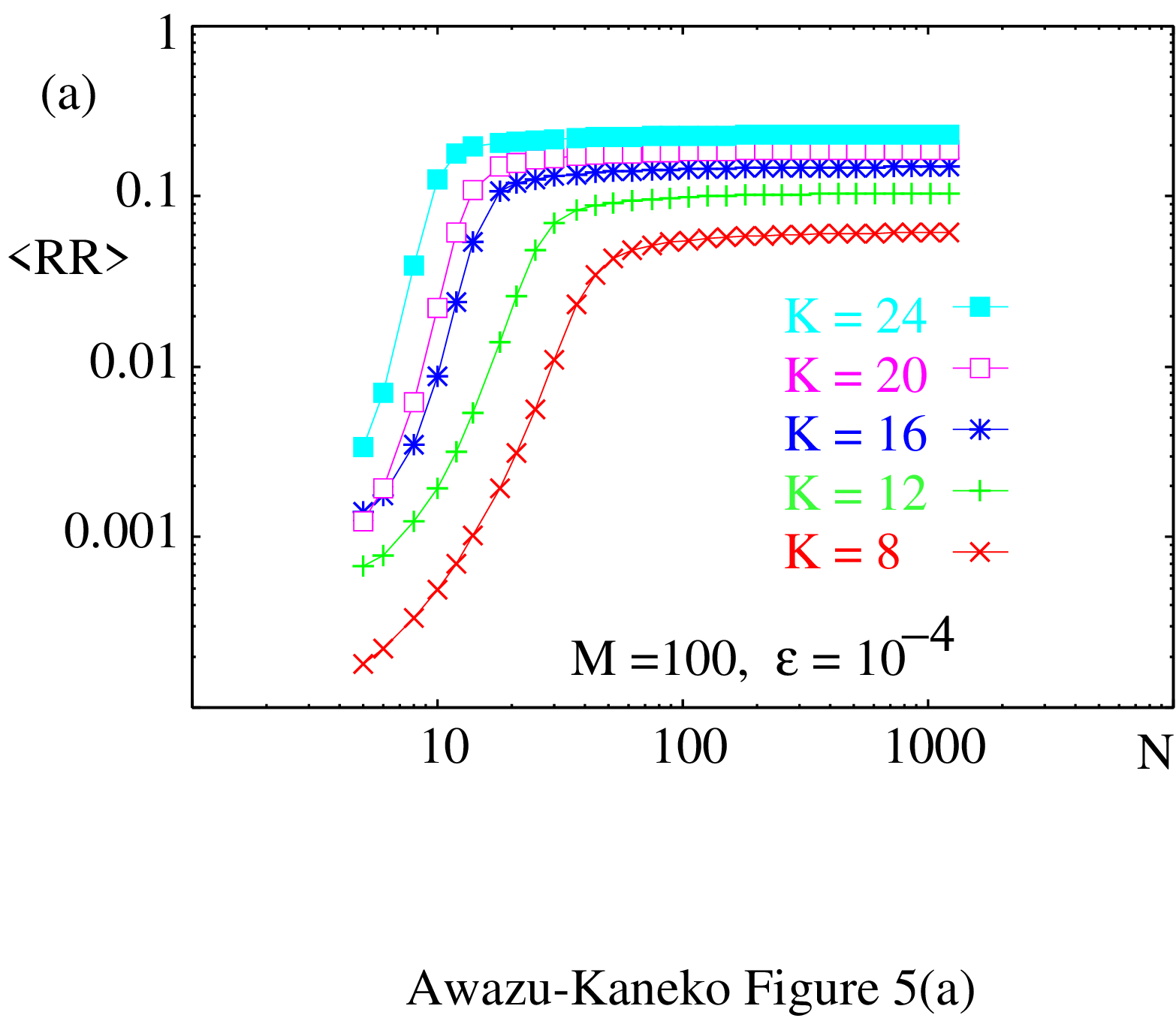}
\includegraphics[width=7.0cm]{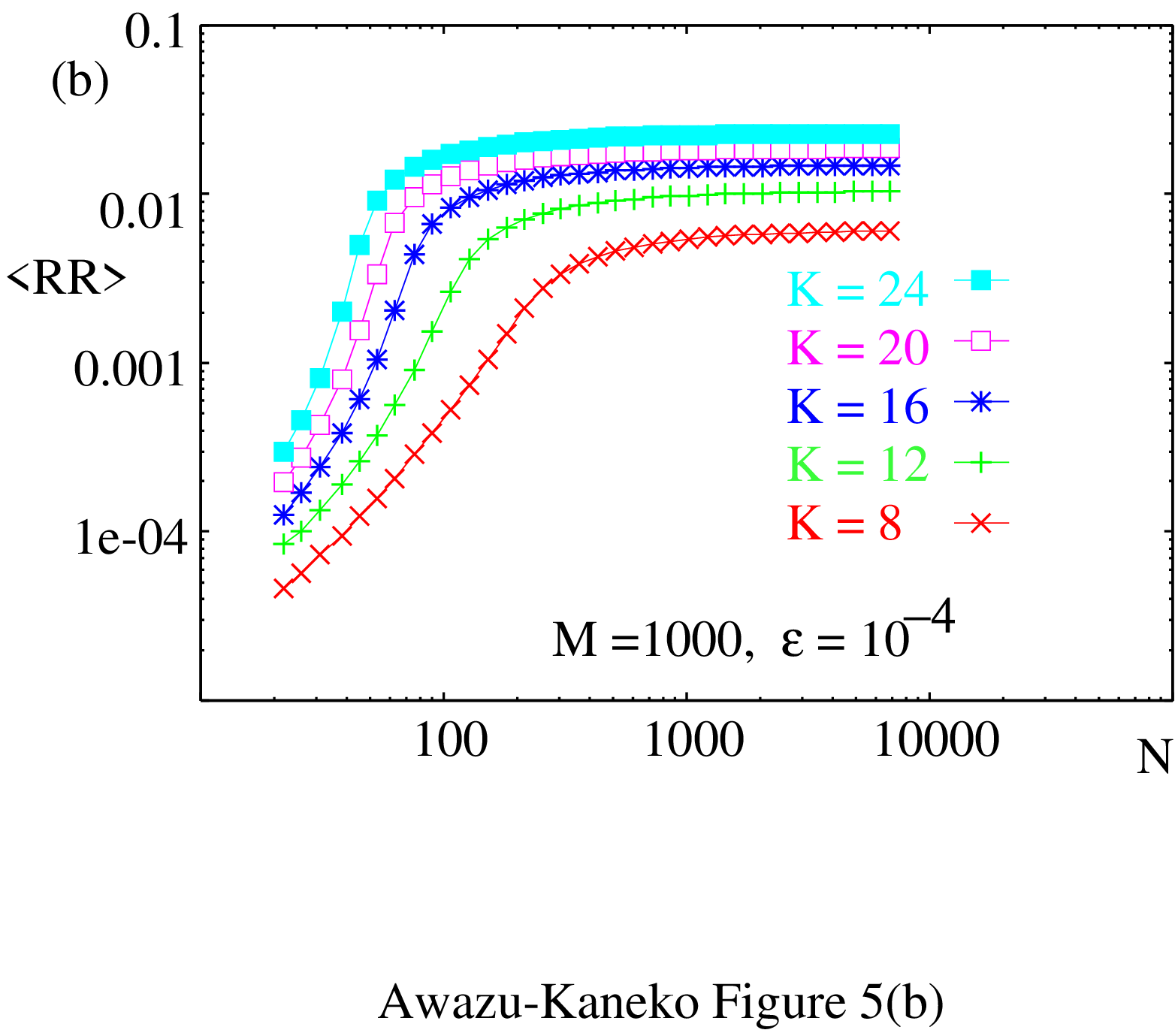}
\end{center}
\caption{Typical examples of $<RR>$ plotted 
as a function of $N$ for several $K$ and 
(a) $M=100$ with $\epsilon = 10^{-4}$ and (b) $M=1000$ with 
$\epsilon = 10^{-4}$. 
} 
\end{figure}

\begin{figure}
\begin{center}
%\psbox[width=8.0cm]{}
\includegraphics[width=7.0cm]{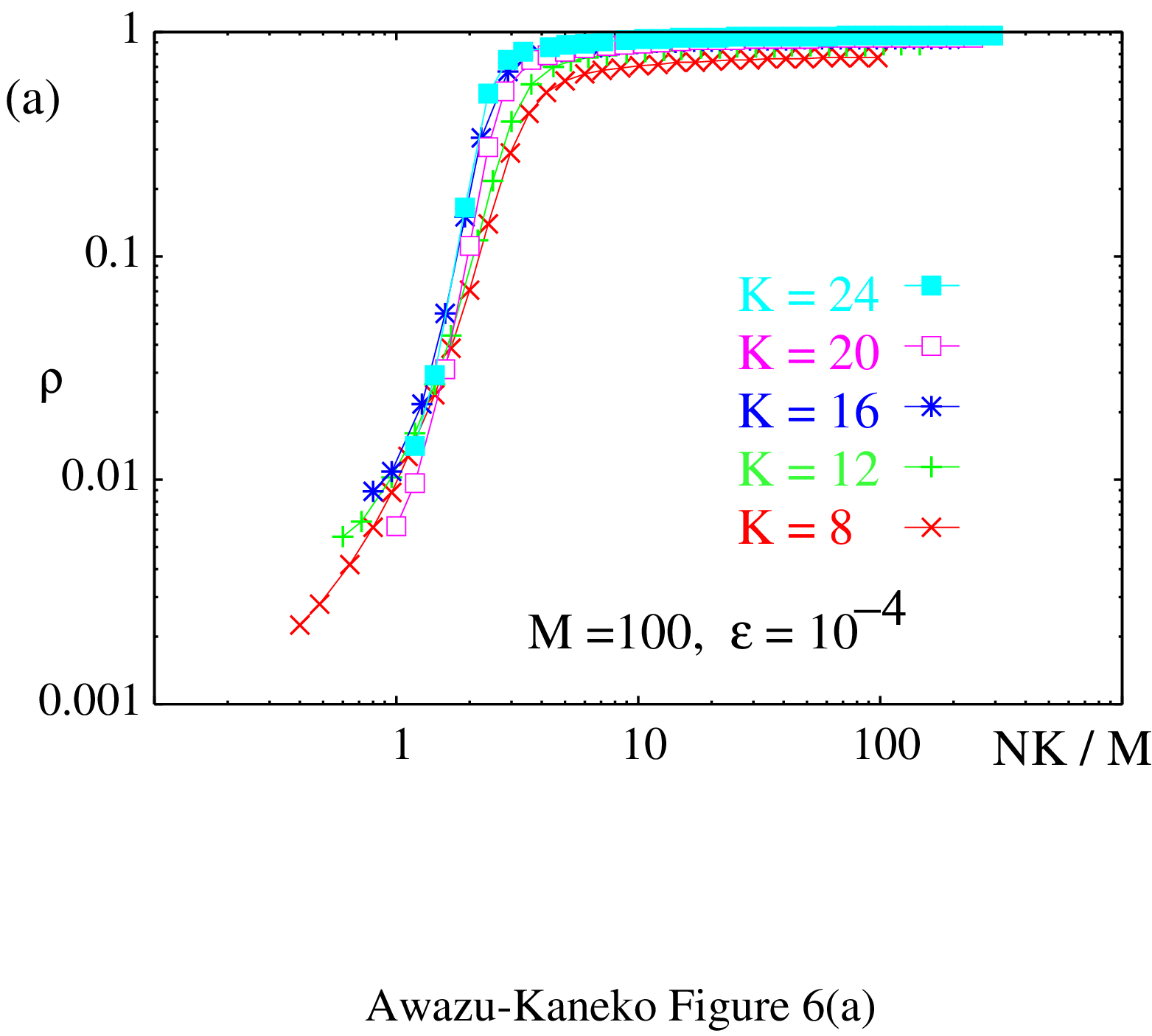}
\includegraphics[width=7.0cm]{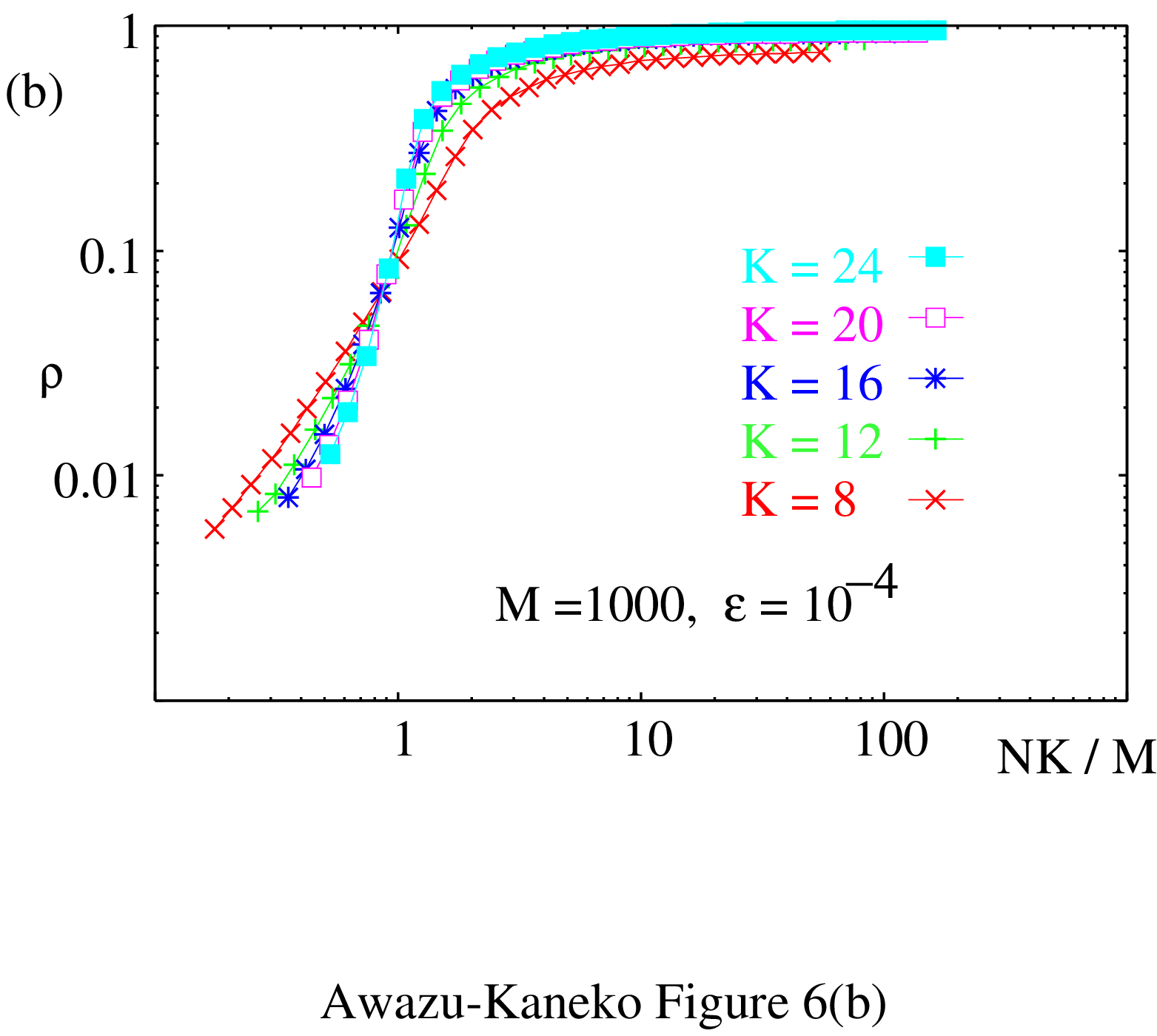}
\includegraphics[width=7.0cm]{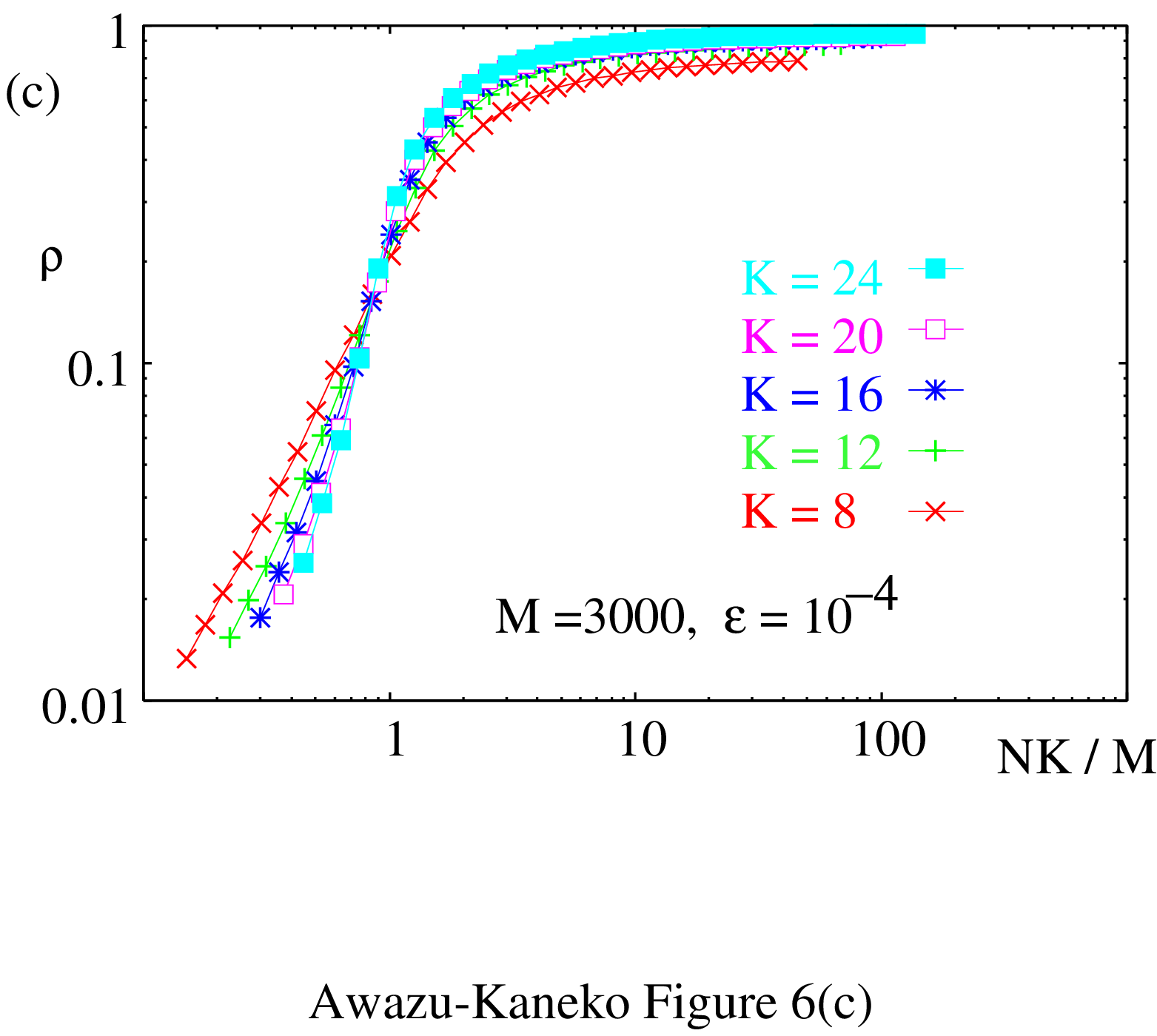}
\end{center}
\caption{Scaled reaction rate $\rho$ as a function of $NK/M$ for
several values of $K$ and (a) $M=100$, (b) $M=1000$ and (c) $M=3000$
with $\epsilon = 10^{-4}$.  }
\end{figure}

\begin{figure}
\begin{center}
%\psbox[width=8.0cm]{}
\includegraphics[width=7.0cm]{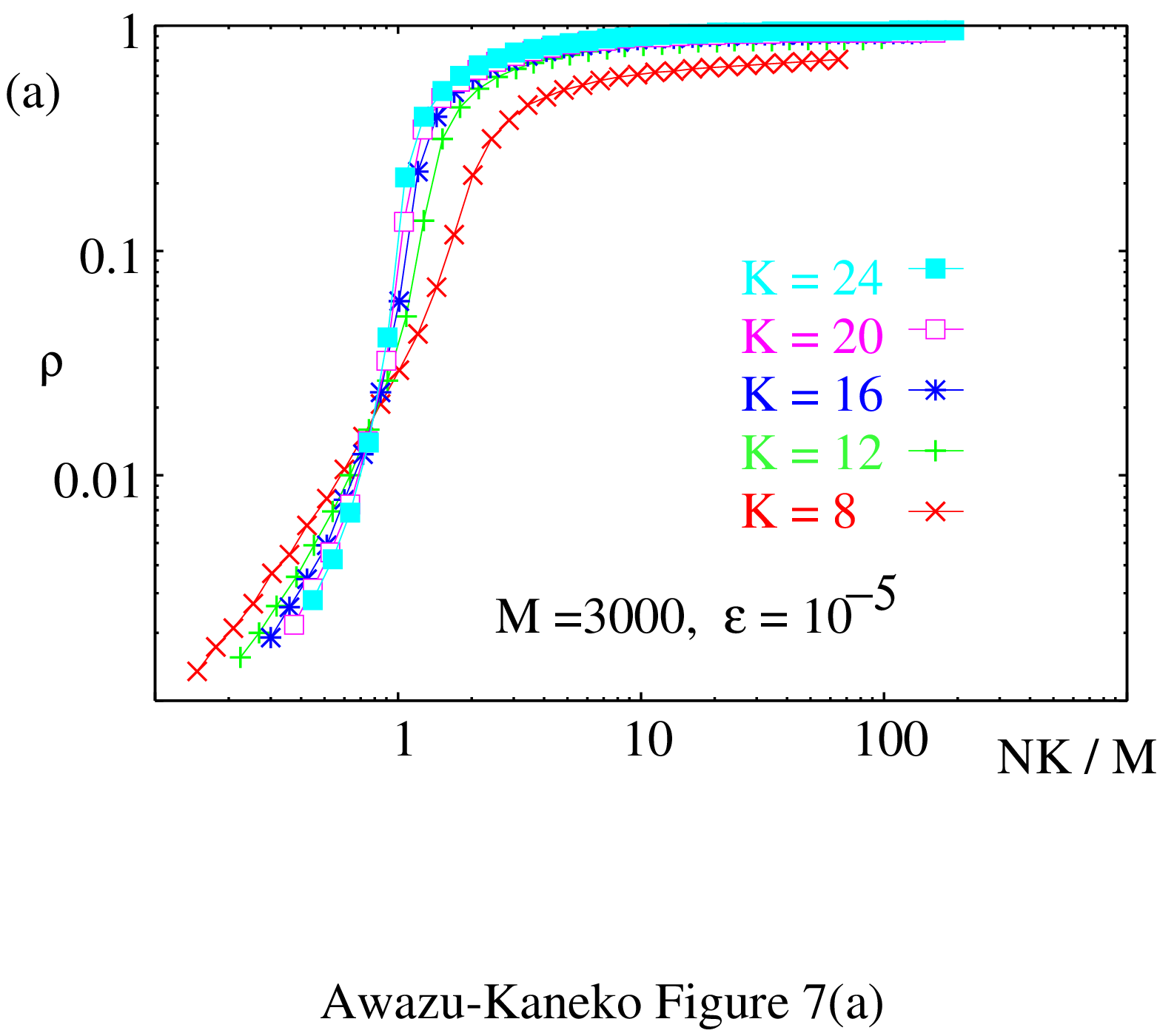}
\includegraphics[width=7.0cm]{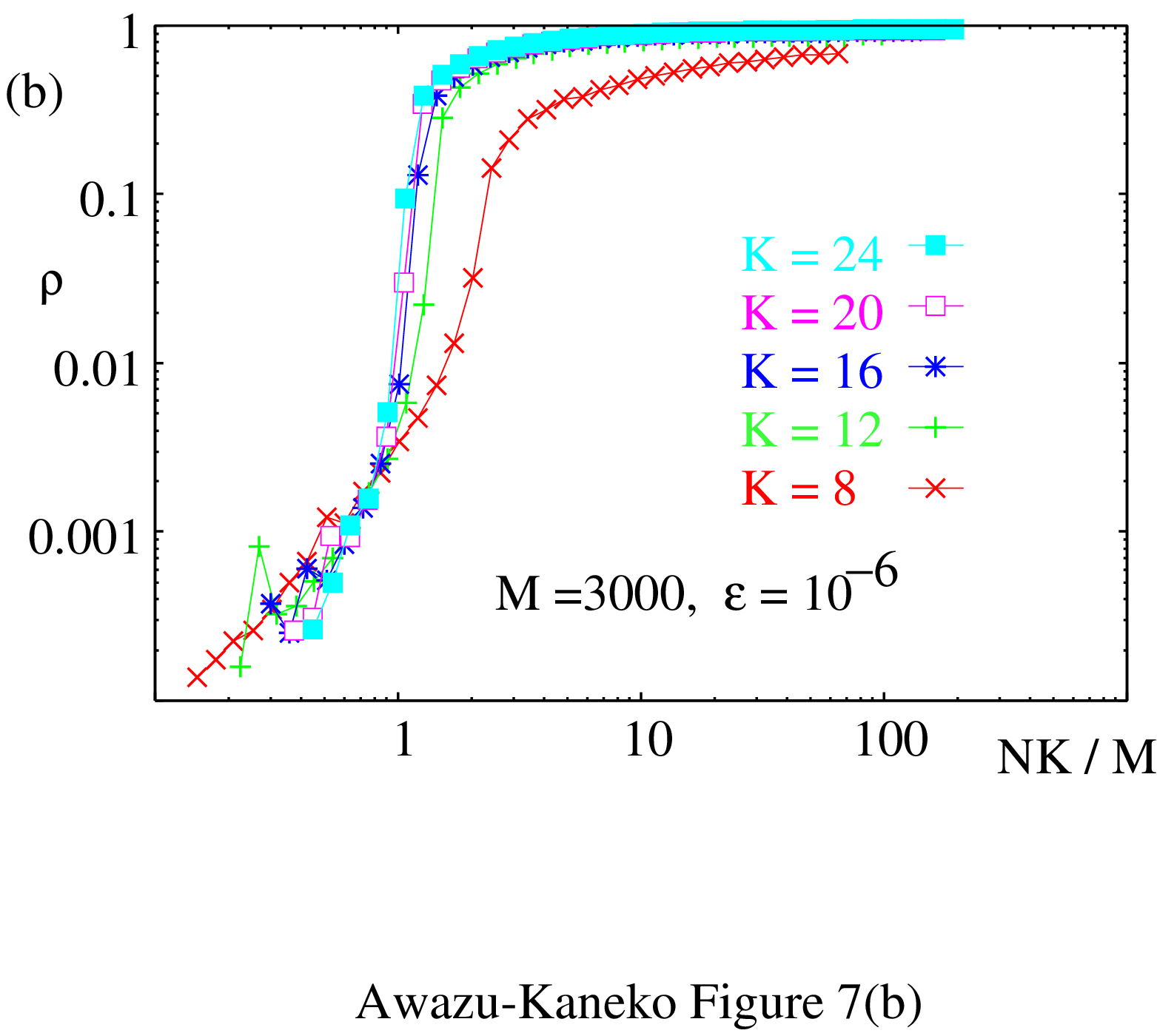}
\end{center}
\caption{Scaled reaction rate $\rho$ as a function of $NK/M$ for several values of $K$ with  
(a) $M=3000$ and $\epsilon = 10^{-5}$ and (b) $M=3000$ and 
$\epsilon = 10^{-6}$. 
} 
\end{figure}


\begin{thebibliography}{9}

\bibitem{cell1} B. Alberts, et. al., Molecular Biology of the Cell 4th
ed. (Garland Science, New York, 2002).

\bibitem{cell2}
N. Olsson,, E. Piek, P. ten Dijke and G. Nilsson, J. Leuko. Biol. 
{\bf 67} (2000) 350.

\bibitem{cell3}
P. Guptasarma, BioEssays  {\bf 17} (1995) 987.

\bibitem{cell4}
H. H. McAdams and A. Arkin, Trends Genet. {\bf 15} (1999) 65.

\bibitem{sto1}
K. Matsumoto and I. Tsuda, J. Stat. Phys. {\bf 31} (1983) 87.

\bibitem{sto2}
W. Horsthemke and R. Lefever, Noise-induced-transitions, ed. H. Haken
	(Springer, Heidelberg, 1984)

\bibitem{sto3}
K. Wiesenfeld and F. Moss, Nature {\bf 373} (1995) 33.


\bibitem{togashi1}
Y. Togashi and K. Kaneko, Phys. Rev. Lett. {\bf 86} (2001) 2459.

\bibitem{togashi2}
Y. Togashi and K. Kaneko, J. Phys. Soc. Jpn. {\bf 72} (2003) 62.


\bibitem{mif1}
B. Hess and A. S. Mikhailov, Science {\bf 264} (1994) 223;
	J. Theor. Biol {\bf 176} (1995) 181. 

\bibitem{mif2}
P. Stange, A. S. Mikhailov and B. Hess, J. Phys. Chem {\bf B 104} (2000) 1844. 

\bibitem{Solomon}
N. M. Shnerb, Y. Louzoun, E. Bettelheim, and S. Solomon,
	Proc. Nat. Acad. Sci.  {\bf 97} (2000) 10322.

\bibitem{togashi3}
Y. Togashi and K. Kaneko, Physica {\bf D 205} (2005) 87. 

\bibitem{marion}
G. Marion, X. Mao, E. Renshaw and J. Liu, Phys. Rev. {\bf E 66} (2002) 051915

\bibitem{zhdanov}
V. P. Zhdanov, Eur. Phys. J. {\bf B 29} (2002) 485.

\bibitem{Kauffman}
S. A. Kauffman, The Origin of Order, Oxford Univ. Press. 1993.

\bibitem{Kaneko-Adv}
K. Kaneko, Adv. Chem. Phys. {\bf 130} (2005) 543.

\bibitem{Furusawa}
C. Furusawa and K. Kaneko, Phys. Rev. Lett. {\bf 90} (2003) 088102.

\bibitem{Ito}
C. Furusawa, et. al., BIOPHYSICS, 1(2005) 25

\bibitem{gene}
G. Lahav, et. al., Nature Genetics. {\bf 36} (2004) 147.

\bibitem{togashi4}
Y. Togashi and K. Kaneko, J. Phys. Cond. Matt.  {\bf 19} (2007) 065150.

\bibitem{graph}
R. Diestel, Graph Theory (Springer-Verlag New York) 

\bibitem{sig}
U. Alon, MG. Surette, N. Barkai and  S. Leibler, Nature {\bf 397}
	(1999) 6715. 

\bibitem{meta}
E. Fisher and U. Sauer, J. Bio. Chem.  {\bf 278} (2003) 46446.

\bibitem{Laha}
G. Lahav, N. Rosenfeld, A. Sigal, N. Geva-Zatorsky, A. J. Levine,
	M. B. Elowitz and U. Alon, Nature Gene. {\bf 36} (2004) 147. 

\bibitem{Miha}
I. Mihalcescu, W. H. Hsing and S. Leibler, Nature {\bf 430} (2004) 81. 

\bibitem{Gupt}
P. Guputasarma, BioEsseys  {\bf 17} (1995) 987. 

%\bibitem{future}
%We will report in the other paper.

\bibitem{net1}
H. Jeong et al. Nature (London) {\bf 407} (2000) 651.

\bibitem{net2}
H. Jeong, M. P. Mason, and A-L. Barabasi, Nature (London) {\bf 411} (2001) 41.

\end{thebibliography}
\end{document}